\documentclass[12pt,onecolumn]{IEEEtran}
\usepackage{amsmath,amsthm}
\usepackage{amsfonts}
\usepackage{paralist}
\usepackage{graphicx}
\usepackage{subfigure}
\usepackage[pdftex]{lscape}

\title{Adaptive selective sidelobe canceller beamformer with applications in radio astronomy}%
\author{Ronny Levanda\footnote{School of Engineering, Bar-Ilan University, 52900, Ramat-Gan, Israel.} \footnote{Email: ronny.levanda@gmail.com} and Amir Leshem }
\date{Aug 22, 2010}%

\begin{document}

\maketitle

\begin{abstract}
We propose a new algorithm, for parameter estimation that is applicable to imaging using moving and synthetic aperture arrays.
 The new method results in higher resolution and more accurate estimation than commonly used methods when strong interfering sources are present inside and outside the
 field of view (terrestrial interference, confusing sources).
\end{abstract}

\section{Introduction}

When multiple antennas or other sensors are used to estimate
incoming signals, it is best to treat them all as a single array and
apply one of the known array processing algorithms for the
estimation. In various situations, however, this is not practical or
even impossible.

In radio-astronomy, the antenna array moves with the rotation of the
Earth. Correlations between the antenna can be obtained for a given
time, but not between different points  in time. Denote by ${\bf
R}_k$ the measured correlation matrix (visibility) of the array at
time $t_k$; the combined correlation matrix of $K$ epochs
$t_{1},\ldots,t_{K}$ can be written as
\begin{equation}
\left(
  \begin{array}{ccc}
    {\bf R}_1 & \emptyset   & \emptyset \\
      \emptyset  & \ddots & \emptyset \\
    \emptyset  & \emptyset  & {\bf R}_K \\
  \end{array}
\right), \label{eq:diag_mat}
\end{equation}
where some of the matrix elements are unknown. The ${\bf R}_k$
matrices need not be of the same size.

The need to overcome this obstacle is as old as the days of radio
astronomy itself. The standard approach to process the measured
correlations (a.k.a visibility) is through using inversion (see \cite{taylor99}  page 128
and \cite{thompson86})
\begin{equation}
\hat{I}(l,m) = \frac{1}{M}  \sum_ {k=1}^M V(u_k,v_k) e^{\frac {2 \pi
j}{ \lambda} (u_k l + u_v m)} \label{eq:classic dirty image}
\end{equation}
where $(u_k,v_k)$ are the baselines (i.e., distances between the
antennas at the measurement time), $M$ is the number of measurements
and $\hat{I}(l,m)$ is the estimated power: in other words,
\emph{dirty image}. This is optimal as long as epochs are
independent. However, sidelobes of interfering sources are not
spatially white, thus making this approach sub-optimal. The dirty
image is used for further processing, by deconvolution algorithms to
yield a better image.

  The most widely used deconvolution algorithm is the CLEAN (proposed by H\"{o}gbom \cite{hogbom74}) and its many variants.
 The CLEAN algorithm assumes that the observed field of view is composed of point sources. CLEAN iteratively removes the brightest point source from the image until the residual image is noise-like. The point sources are accumulated during the iteration and the reconstructed image is the accumulated source list convolved with a reconstruction beam (usually a Gaussian). During the iterations, CLEAN subtracts the strongest point sources either in the image domain or in the visibility domain. The visibility
domain CLEAN is more accurate since we are not limited to pixel
resolution. Throughout this paper, we use the visibility domain
CLEAN.
 Acceleration of the CLEAN algorithm can be achieved by estimating multiple point sources based on a single dirty image (major cycle), as well as defining windows for the search procedure. Practically, defining windows reduces the size of the search space.
 A multi scale CLEAN proposed by Cornwell \cite{cornwell2008a} models the brightness of the sky by the sum of the components of the emission
having different size scales.
 Extensions of the CLEAN algorithm to support  wavelets
 as well as non co-planar arrays are reviewed by Rau et al. in \cite{rao2009}.

A matrix based imaging technique which proved equivalence between
the radio astronomical problem and the signal processing formulation
using beamforming was proposed by Leshem  and van der Veen
\cite{leshem2000a} and further analyzed by Ben-David and Leshem
\cite{bendavid08}. Using the matrix based technique, Leshem and van
der Veen \cite{leshem2000b} suggested a method for the cancellation
of strong interference from a small number of interfering sources.

\subsection{Matrix Based Imaging}
In this section we show the equivalence of the standard approach to
dirty image calculation and classic (i.e., Bartlett) beamforming
(see \cite{levanda2010}, \cite{leshem2004} and \cite{bendavid08} for
an in depth discussion). In this section we assume a calibrated
array (extension for calibration is not hard).

At the $k$'th epoch, the measured correlation matrix is given by
\begin{equation}
 ({\bf R}_k)_{ij}  \equiv
 V(u_{ij}^k,v_{ij}^k)
 \label{eq:corr mat def}
\end{equation}
where $V(u_{ij}^k,v_{ij}^k)$ is the correlation (visibility) between
antenna $i$ and antenna $j$ located at baseline
$(u_{i,j}^k,v_{i,j}^k)$ on the $k$'th epoch. The array steering
vector is defined by
\begin{equation}
\label{eq: classic steering vec}
 {\bf a}_k(l,m) \equiv  \left(
  \begin{array}{c}
    e^{\frac{-2\pi \jmath}{\lambda} (u_{1,0}^k l + v_{1,0}^k m)} \\
    \vdots \\
    e^{\frac{-2\pi \jmath}{\lambda} (u_{P,0}^k l + v_{P,0}^k m)} \\
  \end{array}
\right)
\end{equation}
where $(u_{i,0}^k,v_{i,0}^k)$ is the location of antenna $i$ at the
$k$'th epoch, relative to some convenient point $(u_0,v_0)$, $(l,m)$
are the direction cosines ,$P$ is the number of antennas in the
array and $\lambda$ is the wavelength. For $K$ epochs, we have $M=K
P^2$ and from Equation (\ref{eq:classic dirty image})-(\ref{eq:
classic steering vec}) we obtain that the relation between the dirty
image and the measured correlation matrices is given by
\begin{equation}
\hat{I}(l,m) = \frac{1}{K P^2} \sum_{k=1}^K  {\bf a}_k ^H(l,m)  {\bf
R}_k {\bf a}_k(l,m), \label{eq:classis dirty image textbook}
\end{equation}
which is the classic (Bartlett) beamformer with weight vector ${\bf
w}_k(l,m) = \frac{1}{P} {\bf a}_k(l,m)$.

In the general case, for a weight vector ${\bf w}_k(l,m)$, the dirty
image is given by
\begin{equation}
\hat{I}(l,m) = \frac{1}{K} \sum_{k=1}^K  {\bf w}_k ^H(l,m)  {\bf
R}_k {\bf w}_k(l,m). \label{eq:beamforming dirty image textbook}
\end{equation}

\subsection{MVDR Beamformer}
The MVDR (Minimum Variance Distortionless Response) beam-former is
designed for scenarios that include interfering sources in the field
of view. Its weights are set to minimize the influence of the
interfering sources while passing signals from the desired direction
(i.e., to minimize the interfering power entering the array via its
`sidelobes').

The MVDR is also designed to obey the distortionless response
condition; in the absence of noise, the array input and output
signals should be equal.
  Thus,
  \begin{equation}
 \label{eq:distortionless constraint =1}
{\bf w} ^ H {\bf a} = 1.
 \end{equation}

The MVDR beamformer minimizes the
total array output power. For a given observed source and thermal
noise, the power from interfering sources is minimized. The MVDR
weights are determined by solving the following problem:
\begin{equation}
\label{eqn:MPDR problem def} \left\{
\begin{array}{ccl}
 {\bf w}& =& \arg { \min_{\bf w} {  {\bf w}^H {\bf R} {\bf w}}} \\
 {\bf w} ^ H {\bf a}& =& 1
\end{array}
\right.
\end{equation}
The solution to Equation (\ref{eqn:MPDR problem def}) is given by
\begin{eqnarray}
\label{eq:MVDR and MPDR weights values} {\bf w}^H_{mvdr} & = &
\frac{ {\bf a}^H {\bf R}^{-1}} {{\bf a}^H {\bf R}^{-1} {\bf a}}
\end{eqnarray}

The MVDR is an \emph{adaptive} method. The weights are determined by
the measured visibility. This is unlike the classical beamformer
that determines the weights according to the observing angle
independent of other radiating sources.

\section{Adaptive Selective Sidelobe Canceller Beamforming}
\label{sec:Adaptiv-Selecti-Sidelob-Cancele-Beamfor}
In this section we present a novel image formation technique.
We begin with a simple example that demonstrates the main idea behind the adaptive-selective-sidelobe-canceller (ASSC) algorithm; for a specific observation direction, the received interference through the sidelobes varies strongly as the array rotates.

For simplicity, consider an East-West linear array with $20$ antennas, $\lambda/2$
 spaced, measuring the correlations every $6$ minutes for a 12-hour period.
  The measured correlation matrix at the $k$'th epoch is ${\bf R}_k$.
  For a specific direction $(l,m)$, the output of the $k$'th beamformer,
 $\hat{I}_k = {\bf w}_k^H {\bf R}_k {\bf w}_k,$ is composed of the
 signal-of-interest (SOI) contribution, interfering sources and the
noise contribution. The contribution of the interfering sources is
determined by their location (and strength) relative to the array
sidelobes. Consider a scenario with a few clusters of sources (see
Figure (\ref{fig:sel reasoning 1}a)). Figure (\ref{fig:sel reasoning
1}b) shows the output power of the $k$'th classic beamformer for a
direction of a point source $S_1$ (marked at Figure (\ref{fig:sel
reasoning 1}a)). Figure (\ref{fig:sel reasoning 1}c) shows the
$k$'th MVDR beamformer for the same direction. From all available
time epochs, only a few epochs yield an estimation close
 to the true point source intensity. The intensity estimation of most epochs is biased due to the interference (received through their sidelobes).
 The output power of the two beamformers (classic and MVDR) towards the empty
 direction $S_2$ (marked at Figure (\ref{fig:sel reasoning 1}a)), is
plotted in Figures (\ref{fig:sel reasoning 1}d) and (\ref{fig:sel
reasoning 1}e) respectively. Only a few time epochs yield a close to
zero intensity estimation (the true intensity); whereas most of the
epochs estimate biased intensity originated from the interference
signal received through their sidelobes.
 The time epoch with the \emph{minimal power for a specific direction} yields the best estimator. Averaging the output power for all epochs will result in a biased and inaccurate estimator.

Note that although the number of reliable estimates per direction is
small, the total number of correlation matrices for the entire
FOV is much larger (it depends on the interference location and the
array geometry).  Figure (\ref{fig:sel reasoning 2}) shows
histograms of the number of directions (pixels) for which a specific
correlation matrix estimated the minimal power (i.e., underwent the smallest interference). Simulation conditions are the same as in
Figure ({\ref{fig:sel reasoning 1}). For the classic beamformer
(Figure \ref{fig:sel reasoning 2}a), out of the $181\times181$
pixels in the image, most time epochs (more than $90\%$) performed
best (i.e.,had minimal interference) for at least $661$
pixels. Over $65\%$ of the time epochs benefited from minimal interference
for $90\%$ of the pixels in the image. As for the MVDR beamformer,
of the $121$ available time epochs, most time epochs (more than
$90\%$) performed best (i.e., had minimal interference) for
at least $507$ pixels out of the $181\times181$ pixels in the image.
Over $45\%$ of the time epochs experienced minimal interference for
$90\%$ of the pixels in the image.

\begin{figure}
\subfigure[Original image] 
{
    \includegraphics[width=0.5\textwidth]{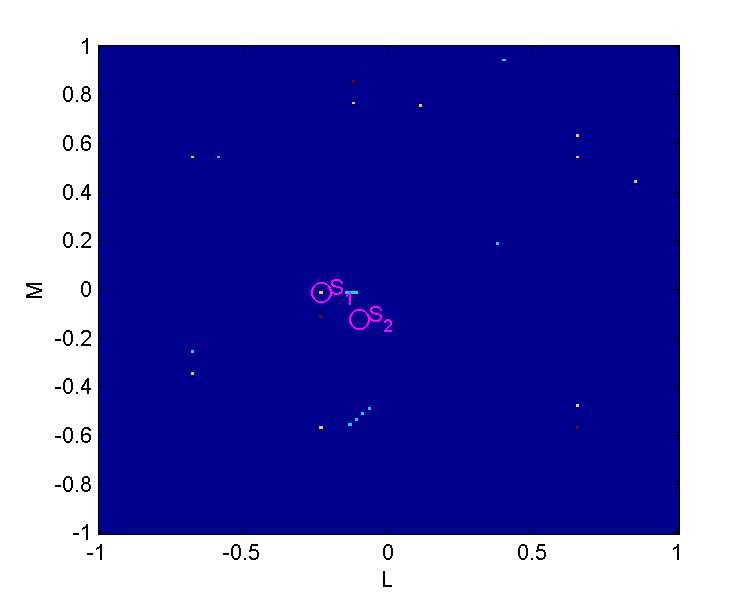}
    } \\
    \subfigure[Classic beamformer outputs for $S_1$ ]
{
    \includegraphics[width=0.5\textwidth]{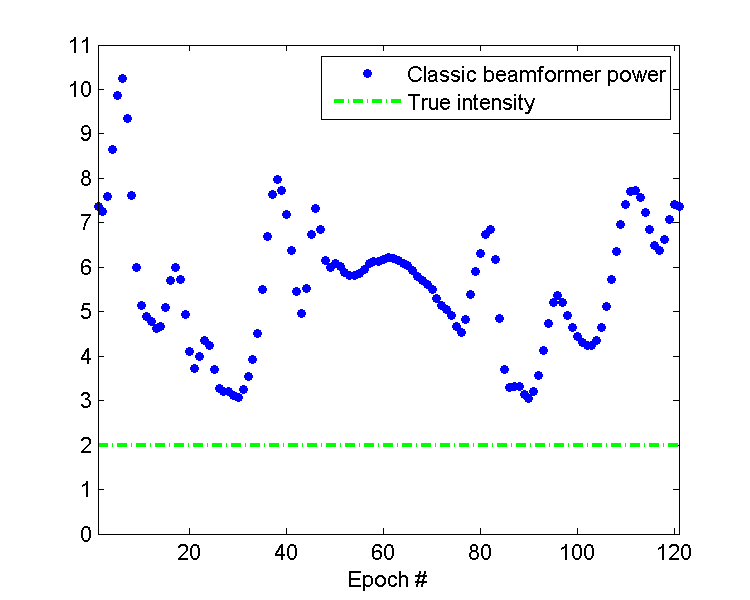}
    } 
    \subfigure[ MVDR beamformer outputs for $S_1$]
{
    \includegraphics[width=0.5\textwidth]{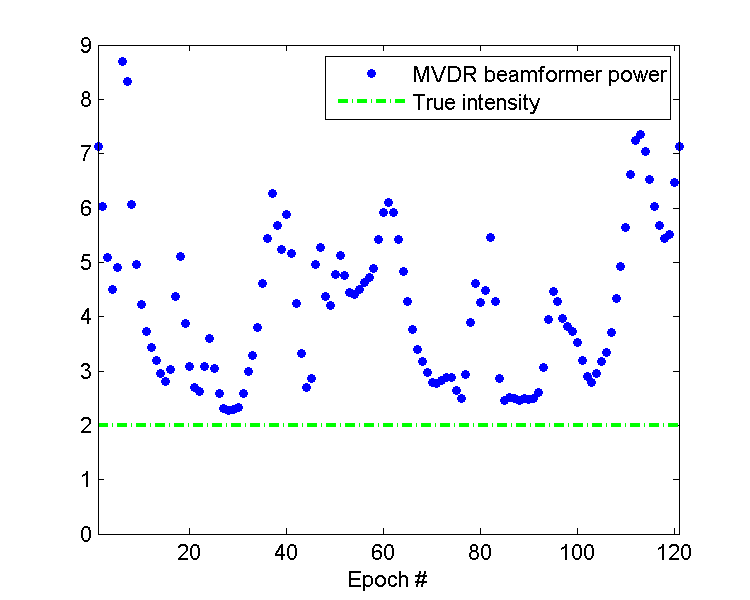}
    } \\
     \subfigure[Classic beamformer outputs for $S_2$ ]
{
    \includegraphics[width=0.5\textwidth]{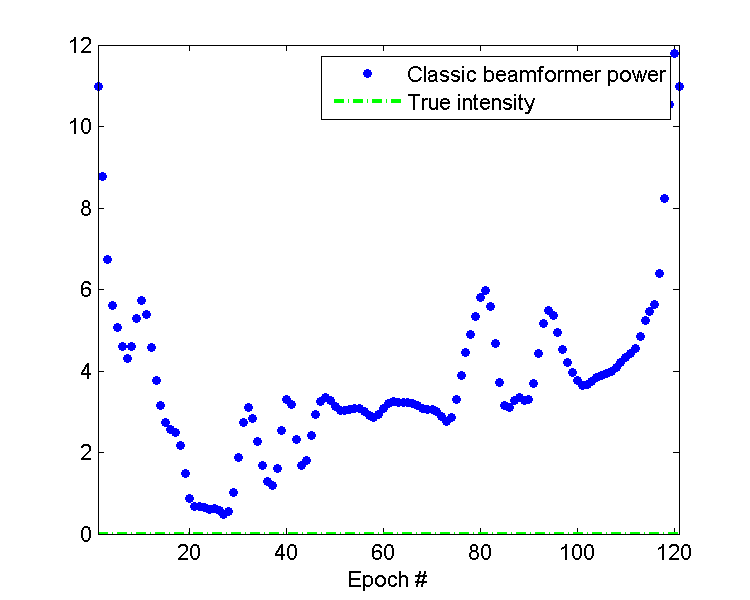}
    } 
     \subfigure[MVDR beamformer outputs for $S_2$  ]
{
    \includegraphics[width=0.5\textwidth]{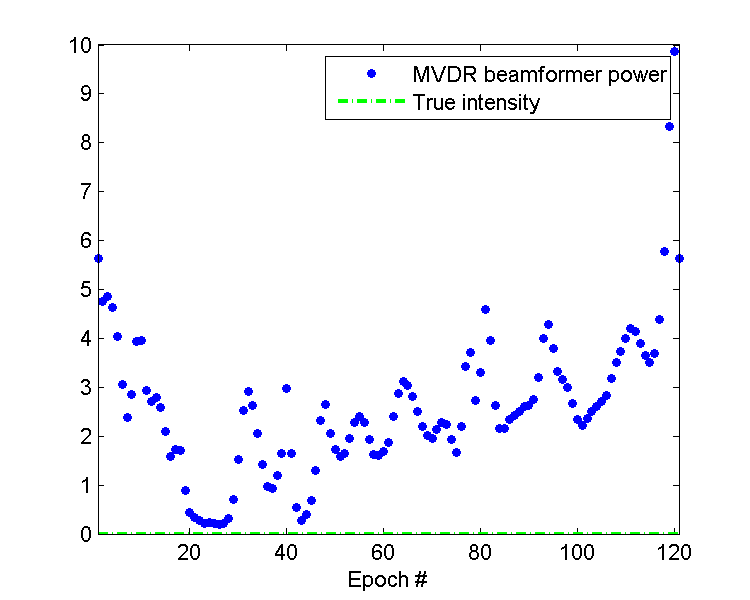}
    } \\
    \caption{The array power for $S_1$ and $S_2$ of the classic and MVDR beamformer for all time epochs.}
\label{fig:sel reasoning 1} 
\end{figure}



\begin{figure}
\subfigure[Classic beamformer]
{
    \includegraphics[width=0.45\textwidth]{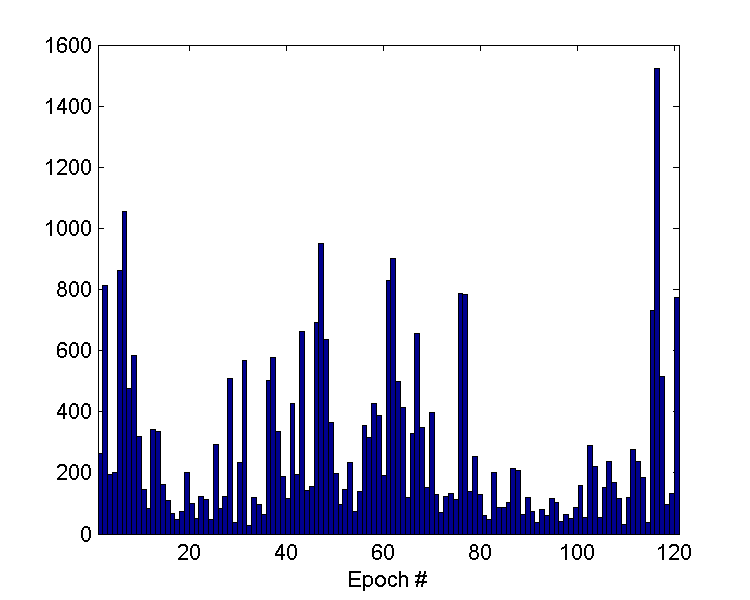}
    } 
    \subfigure[MVDR beamformer]
{
    \includegraphics[width=0.45\textwidth]{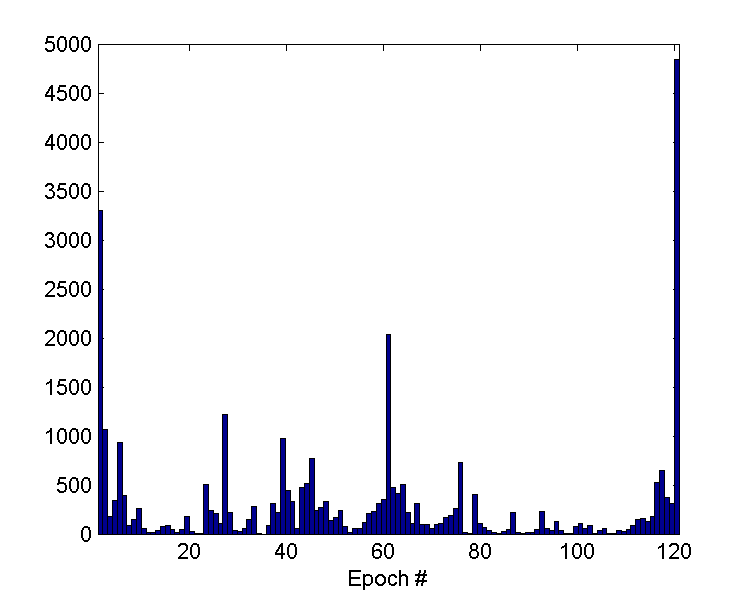}
    } 
    
     \caption{Number of pixels ($(l,m)$ directions), a specific measurement (time epoch) estimated the minimal power (i.e., underwent the smallest interference).}
\label{fig:sel reasoning 2}
\end{figure}

This example demonstrates that the received array power of a specific $(l,m)$ observation direction, varies significantly with the array orientation due to interfering sources. Some of the array orientations yield a reliable intensity estimation, whereas others yield a biased intensity estimation (aggravated by the interfering signals).

\subsection{The ASSC algorithm}
Based on the observations above, we now propose an algorithm that exhibits high performance in the presence of interfering sources and images with a high dynamic range.
For a given set of  $R_k$ correlation matrices, $k=1 \ldots K$, measured at $K$ time epochs (or by $K$ different arrays):
\begin{enumerate}
  \item Calculate the array output power (i.e., dirty image) \emph{for each epoch separately} according to the desired beamformer,
        \begin{equation}
  \hat{I}_k(l,m) =  {\bf w}_k ^H(l,m)  {\bf R}_k {\bf w}_k(l,m).
  \label{eq: dirty image per array}
  \end{equation}

  ${\bf w}_k= {\bf w}_{\texttt{MVDR}}(l,m)$ is the MVDR weight vector given by
 \begin{equation}
 {\bf w}_{\texttt{MVDR}}^H(l,m) = \frac{{\bf a}_k(l,m)^H {\bf R}_k^{-1}}{{\bf a}_k(l,m)^H {\bf R}_k^{-1} {\bf a}_k (l,m)}.
 \label{eq:MVDR weights specific epoch}
 \end{equation}
  \item Determine the ASSC parameters $\tilde{k}$ and $\mu_k$
 where $\tilde{k}$ is the number of best epochs to consider for each $(l,m)$ and $\mu_k$ are their weights. These parameters are best determined using the measured data by plotting a histogram of the calculated ${\hat I}_k(l,m)$ for a specific $(l,m)$. $\tilde{k}$ is determined so no epoch that suffers from significant sidelobes (i.e., has significantly larger power than the minimal power) will be selected. Typically $\tilde{k}< 5\%$ depending on the array geometry and the interference strength and location. As a rule of thumb, the stronger the interference, the smaller the $\tilde{k}$.
   As for $\mu_k$, is should be chosen such that $\mu_{k+1}\leq\mu_k$.
  \item For each $(l,m)$ (each pixel in the image) find the best (i.e., smallest) $\tilde{k}$ values among all measurements,
      \begin{equation}
      \left[ {\check{I}_1(l,m),\ldots,\check{I}_{\tilde{k}}(l,m)} \right]=  [\hat{I}_{(1)},\hat{I}_{(2)},\ldots,\hat{I}_{(\tilde{k})}]
      \end{equation}
      where $ [\hat{I}_{(1)},\hat{I}_{(2)},\ldots,\hat{I}_{(\tilde{k})}]$ are the $\tilde k$ smallest elements in the order statistics of $\left[ {\hat{I}_1(l,m),\ldots,\hat{I}_{K}(l,m)} \right]$
  \item Calculate the ASSC power (dirty image) according to
  \begin{equation}
  \hat{I}^{\texttt{ASSC}}(l,m) = \sum_{k=1}^{\tilde{k}} \mu_k \check{I}_k (l,m).
  \end{equation}
\end{enumerate}
Similarly, the weight vector from Equation (\ref{eq:MVDR weights specific epoch}) can be chosen using any other beamforming technique (for example classic, AAR).
Table (\ref{tab:Many array algo}) summarizes the ASSC beamformer algorithm.

\begin{table}
  \begin{tabular}{|p{1\textwidth}|}
 \hline \\
For each incident angle $(l,m)$:\\
     $\bullet$   Calculate the desired beamformer weight vector for each epoch.\\
     $\bullet$ Calculate the beamformer output power of each correlation matrix separately,  $\hat{I}_k(l,m) = {\bf w}_k^H {\bf R}_k {\bf w}_k$\\
   $\bullet$   Select the best (i.e., smallest) $\tilde{k}$ measurements among $\hat{I}_1(l,m),\ldots \hat{I}_K(l,m)$\\
  $\bullet$    Calculate the ASSC dirty image by $  \hat{I}^{\texttt{ASSC}}(l,m) = \sum_{k = 1}^{\tilde{k}} \mu_k  \hat{I}_{(k)}(l,m)$ \\
    \\     \hline
  \end{tabular}
\caption{ASSC beamforming}
\label{tab:Many array algo}
\end{table}

The computational complexity of the ASSC classic/MVDR beamformer is similar to classic/MVDR beamformers respectively, with the following minor addition: for each pixel, find the $\tilde{k}$ minimal powers from $[\hat{I}_1(l,m),\ldots,\hat{I}_K(l,m)]$.

\subsection{ Rationale for the ASSC}

\begin{figure}
 \includegraphics[width=0.8\textwidth]{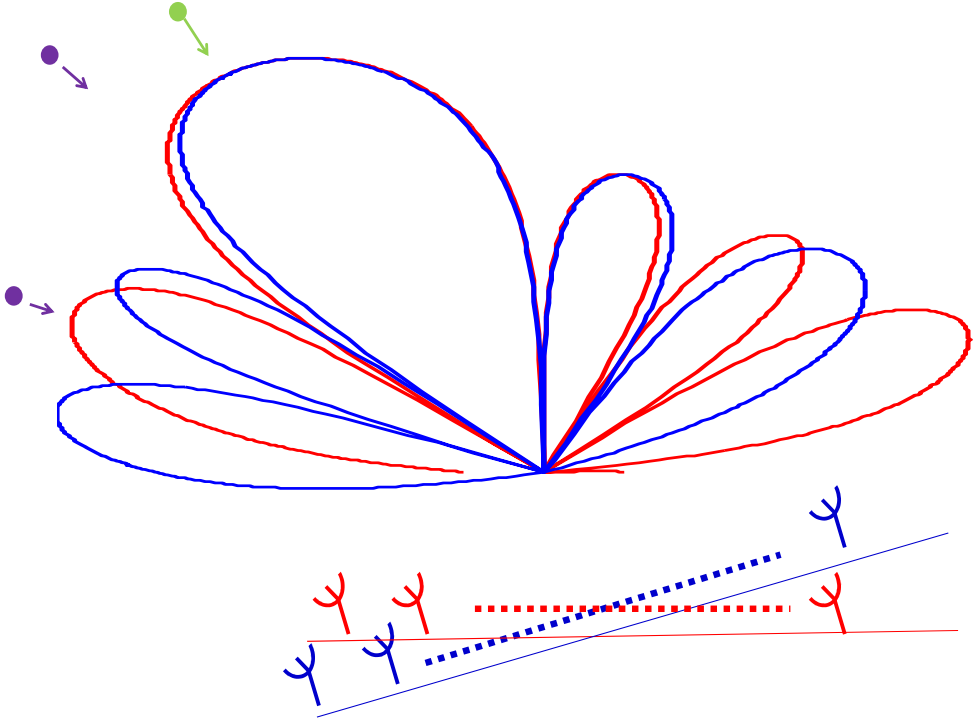}
 \caption{Illustration of array sidelobes.}
 \label{fig:Illustr-array-sidelob}
\end{figure}

Figure \ref{fig:Illustr-array-sidelob} illustrates the sidelobes of a rotating array in two orientations, observing the same SOI (marked in green) in the presence of two interfering sources (marked in purple). The array at the first orientation (marked in red), has a strong sidelobe in the direction of the interfering sources and therefore receives strong interfering power. The array at the second orientation (marked in blue), receives much lower interference power due to the shape and location of its sidelobes relative to the interfering sources.
The received power from the interfering sources depends strongly on the direction of the interfering sources relative to the array sidelobes whereas the received power  from the SOI is similar for all orientations.

The ASSC method is based on the following observations:
\begin{enumerate}[a)]
\item If a signal source is present, and noise at the antenna is neglected, all correlation matrices estimate the same incoming signal and its power.
\item The different results, for different time epochs, are due to interfering signals arriving from other directions through the array sidelobes.
\item By choosing the time epoch with the minimal power, we choose the correlation matrix with the smallest interfering power, which happens to best suppress the interference.
\end{enumerate}

Some comments are in place:
\begin{enumerate}[a)]
\item The proposed method is adaptive and the selected epochs are based on the signal estimates. We implicitly assume that at least one of the array estimations is close to the correct value. If the conditions are such that none of the epochs produce an acceptable estimation, the traditional approaches of averaging may produce more robust results. 
\item In the case where the thermal noise at the antenna is significant and averaging over several measurements is needed, the averaging can be performed on a subset of the epochs with the smallest power.
    \item It should be noted that this technique can be applied to any kind of array beamforming algorithm.
\end{enumerate}

\section{Simulation results}
\label{sec:sim}

 \subsection{In FOV interference}
 This section reports on ASSC algorithm performance compared to existing techniques (classic and MVDR) for the example discussed in (\ref{sec:Adaptiv-Selecti-Sidelob-Cancele-Beamfor}). The ASSC parameters are $\tilde{k} = 3$ and $\mu_k=1$.
 Figure (\ref{fig:twoD example}a) shows the original image that contains a few clusters of sources.
 Using the classic beamformer (Figure (\ref{fig:twoD example}b)), the resulting image (classic dirty image) has wide peaks around each cluster of sources, and the noise is high. The ASSC classic beamformer yield a much quieter image (Figure (\ref{fig:twoD example}c)). The MVDR (Figure (\ref{fig:twoD example}d)) beamformer image has higher spatial resolution than the classic beamformer (as expected). The ASSC MVDR beamformer (Figure (\ref{fig:twoD example}e)) has higher resolution than the MVDR and has the advantage of a quiet image.

\begin{figure}
\subfigure[Original image]
{
    \includegraphics[width=0.5\textwidth]{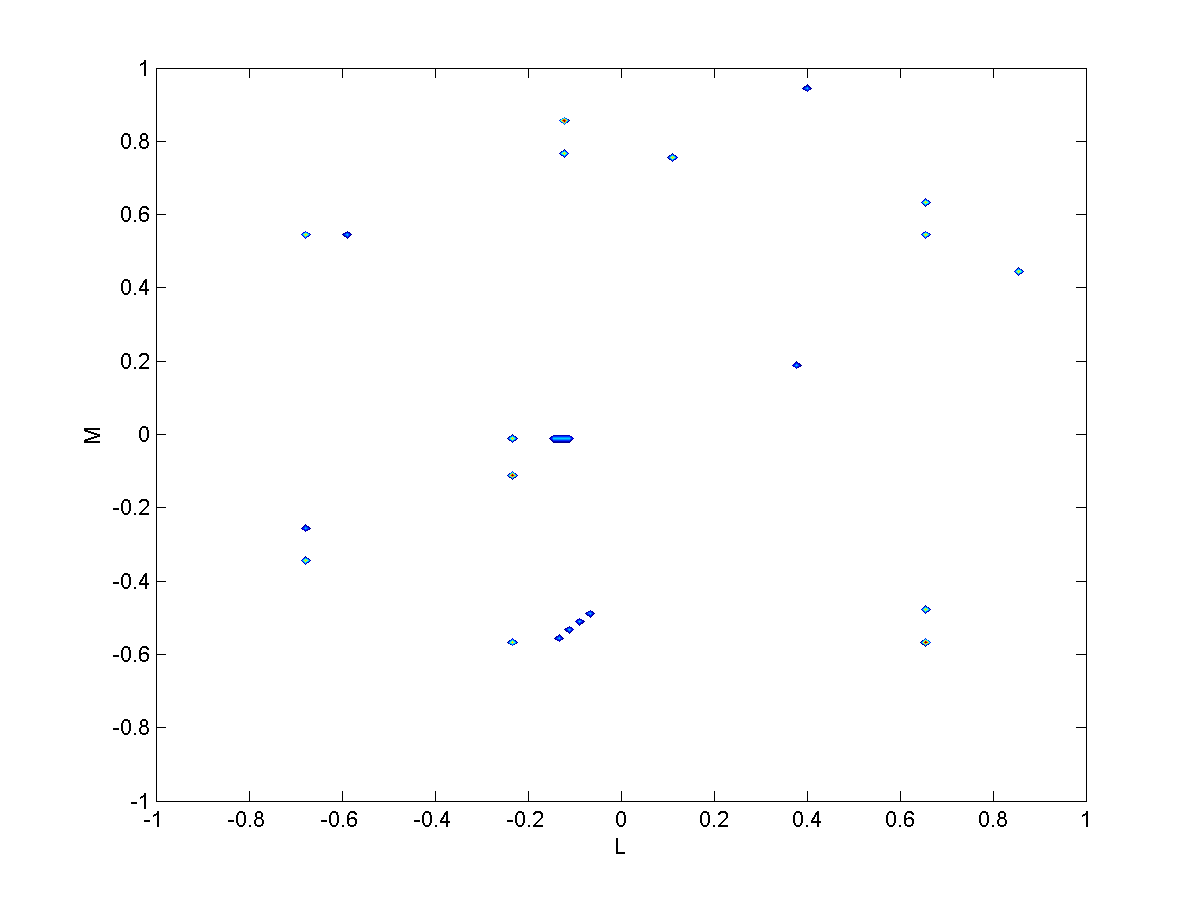}
    } \\
    \subfigure[Classic beamformer]
{
    \includegraphics[width=0.5\textwidth]{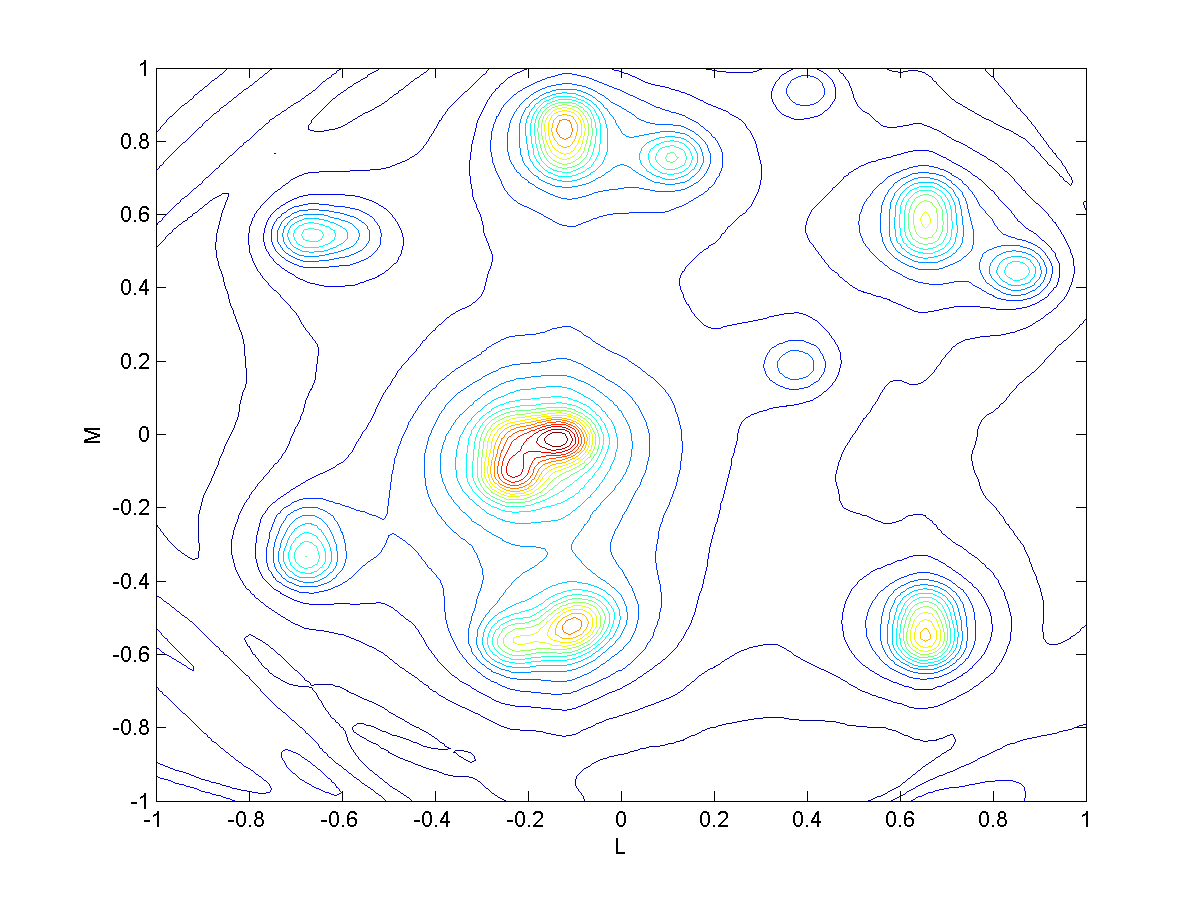}
    }
    \subfigure[ASSC classic beamformer]
{
    \includegraphics[width=0.5\textwidth]{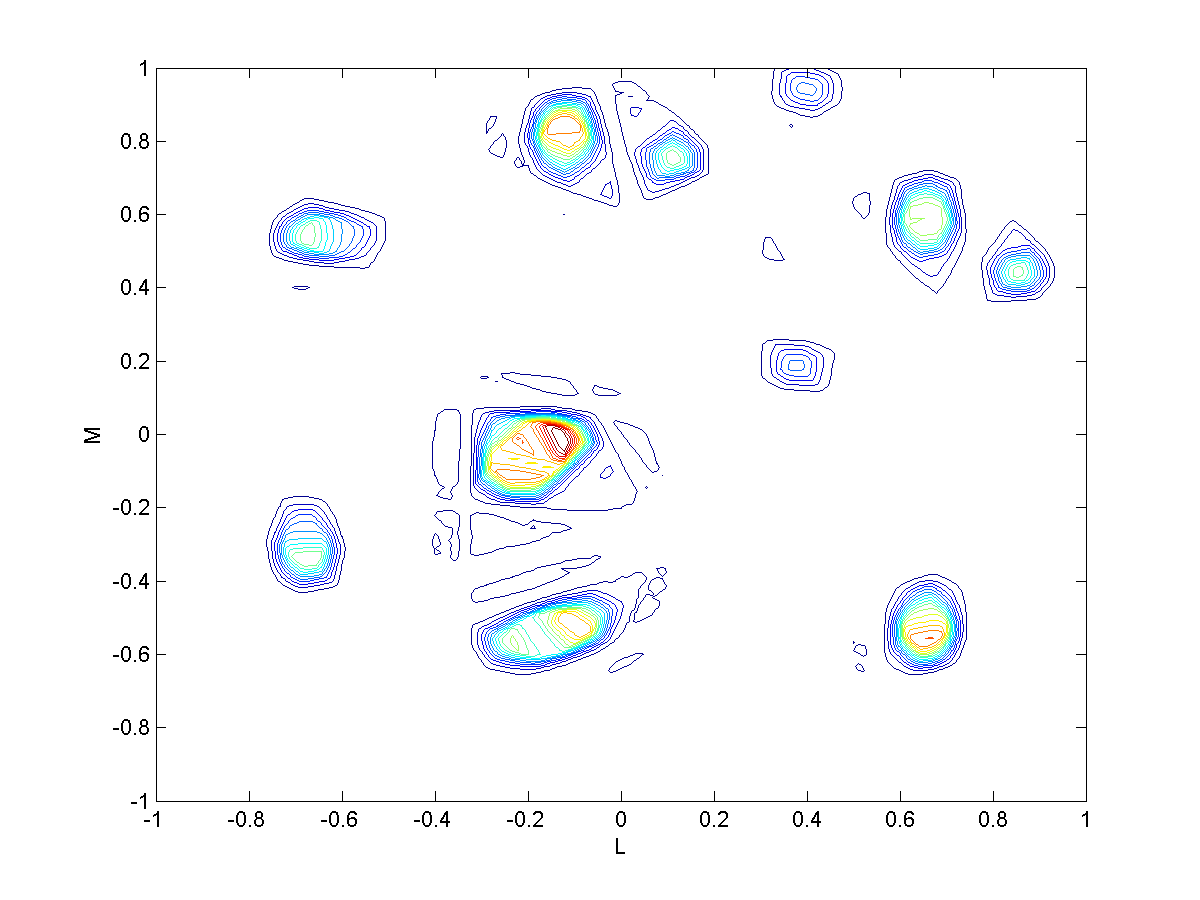}
    } \\
     \subfigure[MVDR beamformer]
{
    \includegraphics[width=0.5\textwidth]{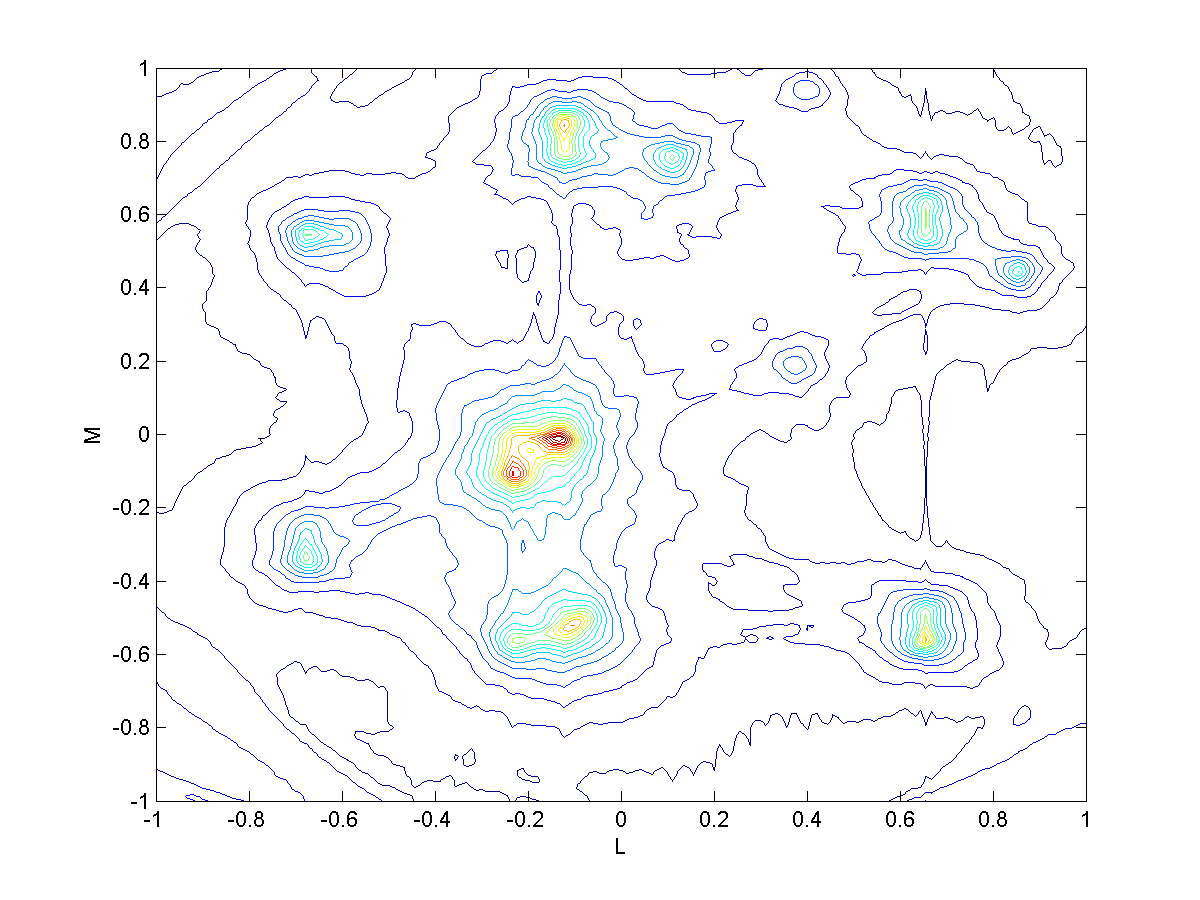}
    }
     \subfigure[ASSC MVDR beamformer]
{
    \includegraphics[width=0.5\textwidth]{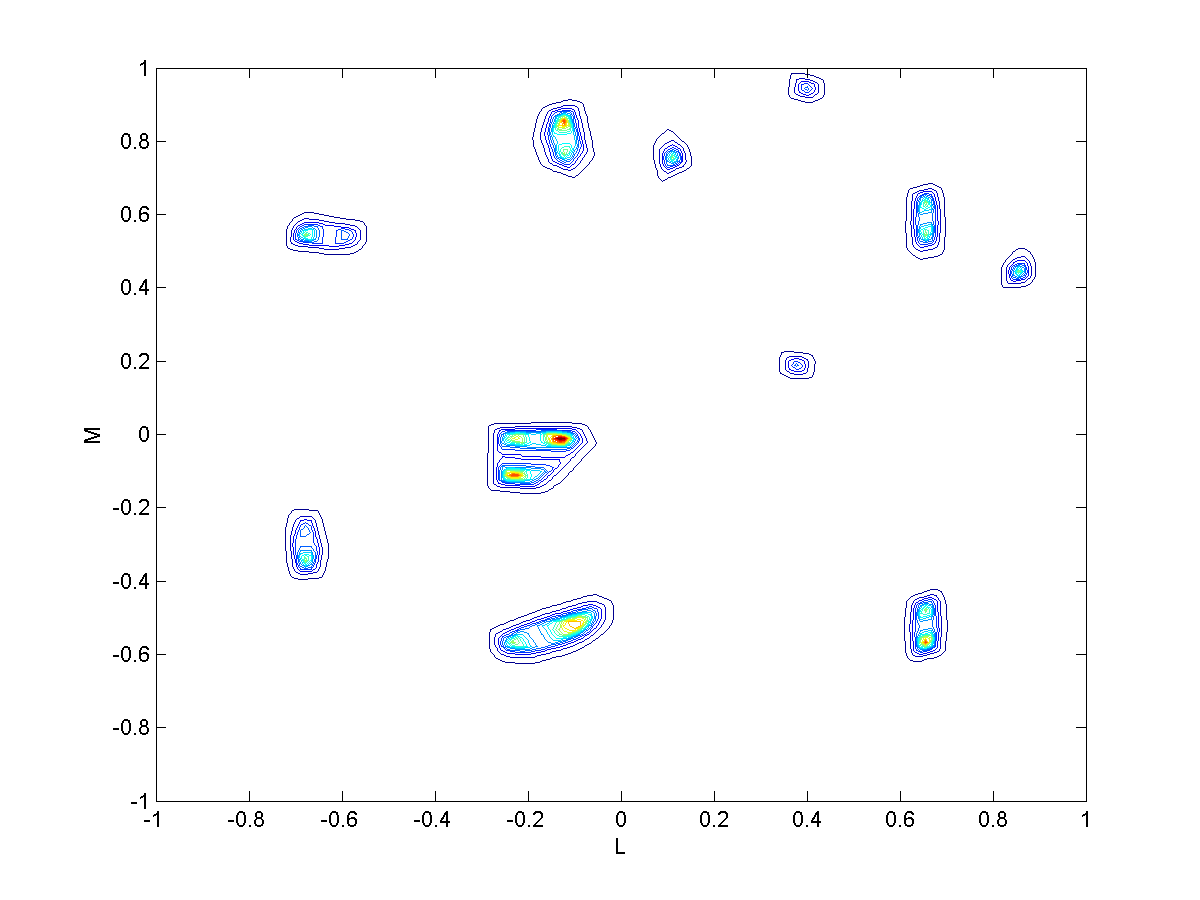}
    } \\
    \caption{Example of dirty images for a few clusters of sources.  All
images are plotted using the same number of contours.}
\label{fig:twoD example}
\end{figure}
    

\subsection{Strong out-of-FOV interference}
This section presents examples of a very strong interference $10^6 $
times the power of the desired sources. Using an East-West array
with $20$ antennas logarithmically spaced $0-200 \lambda$.
 Measurement was done every minute for a 12-hour period. The ASSC parameters used were $\tilde{k}=5$ (out of the $719$ available orientations ) and $\mu_k = \frac{1}{\tilde k}$.
 Figure (\ref{fig:sim strong interference exp}a) shows the original observed image with 6 point sources. The strong interference is not seen in the image (since the interferer is out of the field of view). The output of the classic and MVDR beamformer (dirty images) are shown in Figures (\ref{fig:sim strong interference exp}b) and (\ref{fig:sim strong interference exp}c) respectively. The entire image is smeared with the strong interferer sidelobes. The point sources are not seen. The output of the ASSC MVDR beamformer is shown in Figure (\ref{fig:sim strong interference exp}d). The point sources are seen clearly and the strong interferer sidelobes do not appear in the image at all since only correlation matrices that are affected negligibly from the sidelobes are selected.

\begin{figure}
\subfigure[Original image]
{
    \includegraphics[width=0.5\textwidth]{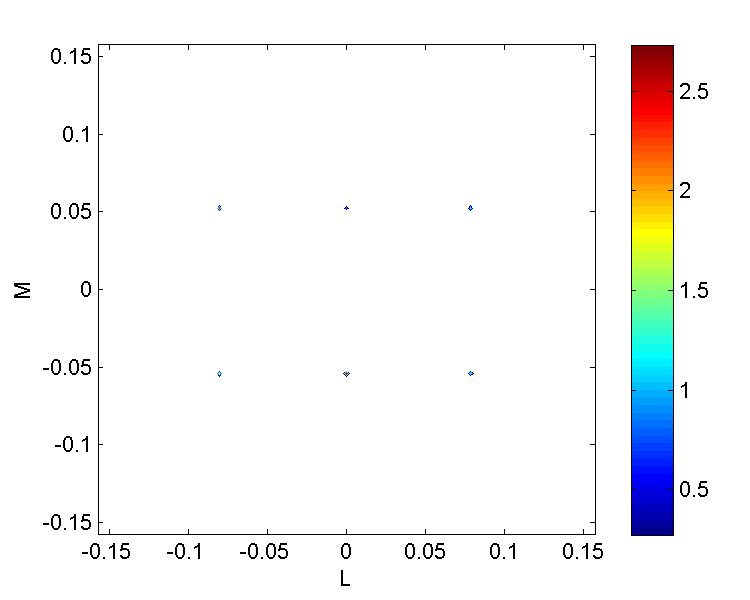}
    } 
    \subfigure[Classic beamformer]
{
    \includegraphics[width=0.5\textwidth]{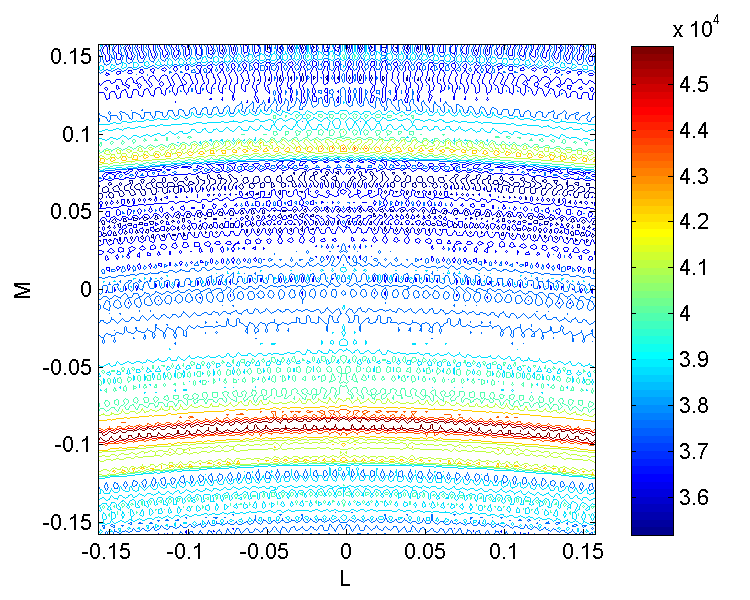}
    }\\
     \subfigure[MVDR beamformer]
{
    \includegraphics[width=0.5\textwidth]{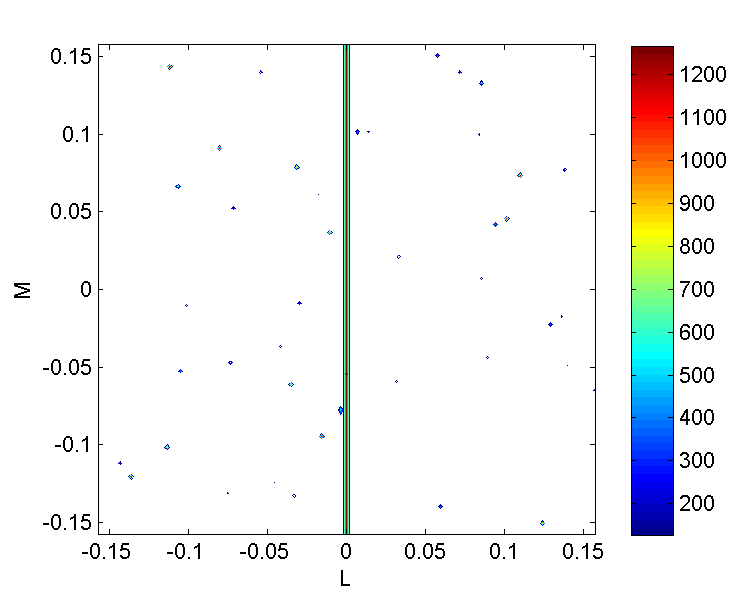}
    }
     \subfigure[ASSC MVDR beamformer]
{
    \includegraphics[width=0.5\textwidth]{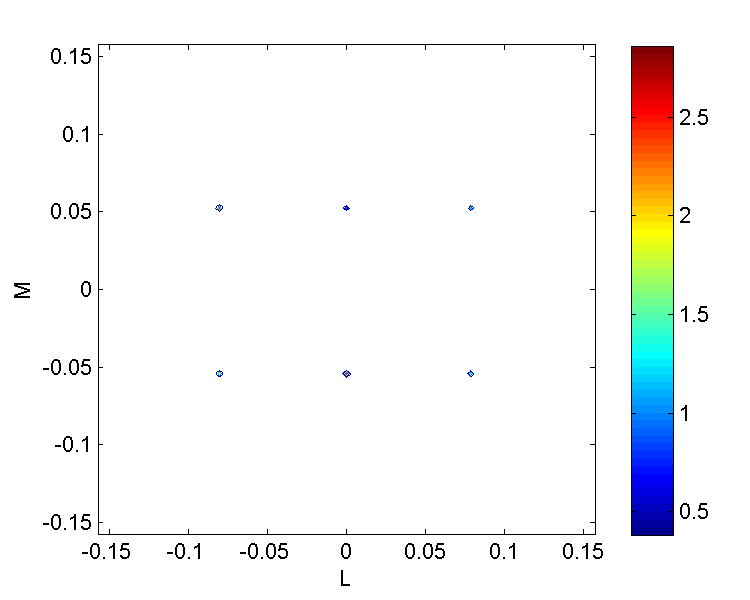}
    } \\
    \caption{Dirty images of 6 point sources and a strong interfering
source outside of the field of view with an intensity $10^6$ times
larger than the sources in the observed field of view.}
\label{fig:sim strong interference exp}
\end{figure}


\section{Summary}
In this paper we introduced the ASSC beamformer, a method to combine
rotating/many array measurements for interference suppression. The performance of the ASSC (classic and MVDR) were demonstrated and compared to the classic and MVDR beamformer. For interference dominant cases, the ASSC beamformer obtains images with higher spatial resolution and interference cancelation  than either the classic or the MVDR beamformer.


\begin{thebibliography}{10}

\bibitem{taylor99}
G.~Taylor, C.~Carilli, and R.~Perley, {\em Synthesis Imaging in
  Radio-Astronomy}.
\newblock Astronomical Society of the Pacific, 1999.

\bibitem{thompson86}
A.~Thompson, J.~Moran, and G.~Swenson, eds., {\em Interferometry and
Synthesis
  in Radio astronomy}.
\newblock John Wiley and Sons, 1986.

\bibitem{hogbom74}
J.~A. H\"{o}gbom, ``Aperture synthesis with nonregular distribution
of
  intereferometer baselines,'' {\em Astron. Astrophys. Suppl}, vol.~15,
  pp.~417--426, 1974.

\bibitem{cornwell2008a}
T.~Cornwell, ``Multiscale {CLEAN} deconvolution of radio synthesis
images,''
  {\em IEEE Journal of Selected Topics in Signal Processing}, vol.~2,
  pp.~793--801, Oct. 2008.

\bibitem{rao2009}
U.~Rau, S.~Bhatnagar, M.~Voronkov, and T.~Cornwell, ``Advances in
calibration
  and imaging techniques in radio interferometry,'' {\em Proceeding of the
  IEEE}, vol.~97, pp.~1472--1481, Aug 2009.

\bibitem{leshem2000a}
A.~Leshem and A.~van~der Veen, ``Radio-astronomical imaging in the
presence of
  strong radio interference,'' {\em IEEE Trans. on Information Theory, Special
  issue on information theoretic imaging}, pp.~1730--1747, August 2000.

\bibitem{bendavid08}
C.~Ben-David and A.~Leshem, ``Parametric high resolution techniques
for radio
  astronomical imaging,'' {\em IEEE Journal of Selected Topics in Signal
  Processing}, vol.~2, pp.~670--684, Oct. 2008.

\bibitem{leshem2000b}
A.~Leshem, A.~van~der Veen, and A.~J. Boonstra, ``Multichannel
interference
  mitigation techniques in radio-astronomy,'' {\em The Astrophysical Journal
  Supplements}, pp.~355--373, November 2000.

\bibitem{levanda2010}
R.~{Levanda} and A.~{Leshem}, ``{Synthetic aperture radio
telescopes},'' {\em
  IEEE Signal Processing Magazine}, vol.~27, pp.~14--29, Jan. 2010.

\bibitem{leshem2004}
A.~van~der Veen, A.~Leshem, and A.~Boonstra, ``Array signal
processing in
  radio-astronomy,'' {\em Experimental Astronomy}, vol.~17, pp.~231--249, June
  2004.

\bibitem{capon69}
J.~Capon, ``High resolution frequency-wavenumber spectrum
analysis,'' {\em
  Proceedings of the IEEE}, pp.~1408--1418, 1969.

\bibitem{vantrees}
H.~V. Trees, {\em Optimum array processing}.
\newblock J. Wiley, 2002.
%
%
\end{thebibliography}


\end{document}